\documentclass[11pt]{article}

\begin{document}
\renewcommand{\baselinestretch}{1.3}

\title{NEW PHYSICS FROM A DYNAMICAL VOLUME ELEMENT}

\author{Eduardo Guendelman and Alexander Kaganovich\\
\small\it Department of Physics, Ben-Gurion University, Beer-Sheva, Israel  \\[-1.mm]
\small\it email: guendel@bgumail.bgu.ac.il, alexk@bgumail.bgu.ac.il \\
${}$ \\
Emil Nissimov and Svetlana Pacheva\\
\small\it Institute for Nuclear Research and Nuclear Energy,\\[-1.mm]
\small\it Bulgarian Academy of Sciences, Sofia, Bulgaria  \\[-1.mm]
\small\it email: nissimov@inrne.bas.bg, svetlana@inrne.bas.bg}

\maketitle

\begin{abstract}
The use in the action integral of a volume element of the form
$\Phi d^{D}x$ where $\Phi$ is a metric independent measure can
give new interesting results in all types of known generally
coordinate invariant theories: (1) 4-D theories of gravity plus
matter fields; (2) Reparametrization invariant theories of extended
objects; (3) Higher dimensional theories. In the case (1), a large
number of new effects appears: under normal particle physics
conditions (primordial) fermions split into three families; when
matter is highly diluted, neutrinos increase their mass and can
contribute both to dark energy and to dark matter. In the case
(2), it leads to dynamically induced tension; to string models of
non abelian confinement; to the possibility of new Weyl-conformally invariant
light-like branes which dynamically adjust themselves to sit at
black hole horizons; in the context of higher dimensional theories
it can provide examples of massless 4-D particles with nontrivial
Kaluza Klein quantum numbers. In the case  (3), i.e. in brane and
Kaluza Klein scenarios, the use of a metric independent measure
makes it possible to construct naturally models where only the
extra dimensions get curved and the 4-D remain flat.

\end{abstract}

\section{Introduction}

 We have studied models of the new class of
theories\cite{G1}-\cite{G20} based on the idea that the action
integral may contain the new metric-independent measure of
integration. For example, in four dimensions the new measure can
be built of four scalar fields $\varphi^{i}$ \, ($i=1,2,3,4$)
\begin{equation}
\Phi
=\varepsilon^{\mu\nu\alpha\beta}\varepsilon_{ijkl}\partial_{\mu}\varphi^{i}
\partial_{\nu}\varphi^{j}\partial_{\alpha}\varphi^{k}
\partial_{\beta}\varphi^{l}.
\label{Phi}
\end{equation}
$\Phi$ is the scalar density under general coordinate
transformations and the action can be chosen in the form $S = \int
L\Phi d^{4}x$. This has been applied to three different
directions:

\medskip
I. Investigation of the four-dimensional gravity and matter fields
models containing the new measure of integration that appears to
be promising for resolution of the dark energy and dark matter
problems, fermion families problem, the fifth force problem, etc..

\medskip
II. Studying new type of string and brane models based on the use
of a modified world-sheet/world-volume integration measure. It
allows new types of objects and effects like for example:
spontaneously induced string tension; classical mechanism for a
charge confinement; Weyl-conformally invariant light-like (WILL) brane having
the promising results for black hole physics.

\medskip
III. Studying higher dimensional realization of the idea of the
modified measure in the context of the Kaluza-Klein and brane
scenarios with the aim to solve the cosmological constant problem.

\section{Gravity, Particle Physics and Cosmology}.

Since $\Phi$ (\ref{Phi}) is a
total derivative, a shift of $L$ by a constant has no effect on
the equations of motion. Similar shift of $L$ in usual theories,
i.e. with the action $S = \int L\sqrt{-g} d^{4}x$, would lead to
the shift of the constant part of the Lagrangian which in the
Einstein's GR is the cosmological constant. The exploitation of
this circumstance for a resolution of the "old" cosmological
constant problem was the initial motivation\cite{G1} for using the
measure $\Phi$ instead of $\sqrt{-g}$. It turns out that working
with the volume element $\Phi d^{4}x$ {\em instead of} $\sqrt{-g}
d^{4}x$ it is impossible to construct realistic models, e.g. with
a nontrivial scalar field dynamics.

However the situation is dramatically changed if one to apply the
action principle to the action of the general form
\begin{equation}
    S = \int L_{1}\Phi d^{4}x +\int L_{2}\sqrt{-g}d^{4}x,
\label{S}
\end{equation}
 including two Lagrangians $ L_{1}$ and $L_{2}$ and two
measures of the volume elements ($\Phi d^{4}x$ and
$\sqrt{-g}d^{4}x$ respectively). To provide parity conservation,
one can  choose for example one of $\varphi^{i}$'s to be
pseudoscalar. Constructing the field theory with the action
(\ref{S}), we make only two basic additional  assumptions:

(A) $L_{1}$ and $L_{2}$ are independent of the measure fields
$\varphi_{a}$. Then the action (\ref{S}) is invariant under volume
preserving diffeomorphisms\cite{G2},\cite{G6}. Besides, it is
invariant (up to an integral of a total divergence) under the
infinite dimensional group of shifts of the measure fields
$\varphi^{i}$: $\varphi^{i}\rightarrow\varphi^{i}+f^{i}(L_{1})$,
where $f^{i}(L_{1})$ is an arbitrary differentiable function of
the Lagrangian density $L_{1}$. This symmetry prevents the
appearance of terms of the form $h(\Phi/\sqrt{-g})\Phi$ in the
effective action (where quantum corrections are taken into
account) with single possible exception when the function
$h(\Phi/\sqrt{-g})$ is of the form
$h(\Phi/\sqrt{-g})=c\sqrt{-g}/\Phi$; here $c$ is $\Phi/\sqrt{-g}$
independent but may be a function of all other fields. This makes
possible that quantum corrections generate an additive
contribution to the cosmological constant term which may present
also in the second term of the action $(\ref{S})$. Moreover, one
can think of a theory where we start from the action $\int L\Phi
d^{4}x$ but quantum effects modify $L$ to some new Lagrangian
density $L_{1}$ and they also generate a term
$h(\Phi/\sqrt{-g})\Phi$ with
$h(\Phi/\sqrt{-g})=L_{2}\sqrt{-g}/\Phi$. In other words, the
action of the form $(\ref{S})$ may be an effective quantum action
corresponding to the classical action $\int L\Phi d^{4}x$. The
structure of the action (\ref{S}) may be motivated also in the
brane scenario\cite{G12}.

(B) We proceed in the first order formalism where all fields,
including metric $g_{\mu\nu}$ (or vierbeins ${e}_{a\mu}$),
connection coefficients (or spin-connection $\omega_{\mu}^{ab}$)
and the measure fields $\varphi^{i}$ are independent dynamical
variables. All the relations between them follow from equations of
motion. The  field theory based on the listed assumptions we call
"Two Measures Theory" (TMT).

It turns out that the measure fields $\varphi_{a}$ affect the
theory only via the ratio of the two measures
\begin{equation}
\zeta\equiv \Phi /\sqrt{-g} \label{zeta}
\end{equation}
 which is the scalar field. It is determined by a constraint in
the form of an algebraic equation which is exactly a consistency
condition of equations of motion.  {\it The constraint determines
$\zeta$ in terms of the fermion density and scalar fields}.

Applying the Palatini formalism in TMT one can show (see for
example\cite{G6} or Appendix C of Ref.\cite{G12})  that the
resulting relation between metric and  connection includes also
the gradient of $\zeta$. This means that with the original set of
variables we are not in a Riemannian (or Riemann-Cartan)
space-time. Gradient of $\zeta$ presents also in all equations of
motion. By an appropriate change of the dynamical variables which
consists of a conformal transformation of the metric and a
multiplicative redefinitions of the fermion fields, one can
formulate the theory as that in a Riemannian (or Riemann-Cartan)
space-time. The corresponding conformal frame we call "the
Einstein frame". The big advantage of TMT is that in the very wide
class of models, {\em the equations of motion in the Einstein
frame take the canonical general form of those of GR}, including
the field theory models in curved space-time. All the novelty
consists in the structure of the scalar fields effective
potential, masses of fermions and their interactions to scalar
fields as well as the structure of fermion contributions to the
energy-momentum tensor: all these now depend via $\zeta$ on the
scalar fields (e.g., dilaton, Higgs) and the fermion energy
densities. In addition to the canonical fermion contribution to
the energy-momentum tensor there appears the non-canonical one
proportional to $g_{\mu\nu}m(\zeta)\bar{\Psi}\Psi$, where
$m(\zeta)$ is the effective $\zeta$ depending "mass" of the
primordial fermion.

The surprising feature of the theory is that although  the
gravitational equations are used for obtaining the constraint,
neither Newton constant nor curvature appears in the constraint.
This means that the {\it geometrical scalar field} $\zeta (x)$ is
determined by the matter fields configuration locally and
straightforward (that is without gravitational interaction). As a
result of this, $\zeta (x)$ has a decisive influence in the
determination of the effective (that is appearing in the Einstein
frame) interactions and particle masses, and due to this, in the
gravity and particle physics, cosmology and astrophysics.

In Ref.\cite{G9}-\cite{G11} we have started to study the models
with the most general form for $L_1$ and $L_2$ (without higher
derivatives) such that the action (\ref{S}) possesses  both a
non-Abelian gauge symmetry and a special type of scale symmetry
(the latter includes the shift symmetry\cite{Carroll} of the
dilaton $\phi\rightarrow\phi +const$). For short, in a schematic
form $L_1$ can be represented as
\begin{equation}
L_1=e^{\alpha\phi /M_{p}}\left[-\frac{1}{\kappa}R(\omega ,e)
-\frac{1}{2}g^{\mu\nu}\phi_{,\mu}\phi_{,\nu}+(Higgs)+(gauge)+(fermions)\right]
 \label{L1}
\end{equation}
and similarly for $L_2$ (with different choice of the
normalization factors in front of each of the terms). Here
$R(\omega ,e)$ is the scalar curvature in the first order
formalism where the spin-connection $\omega_{\mu}^{ab}$ and the
vierbein $e^{a}_{\mu }$ are independent; $M_P$ is the Planck mass;
$\alpha$ is the dimensionless parameter. Varying the measure
fields $\varphi_{a}$ and assuming $\Phi\neq 0$, we get equations
that yield
\begin{equation}
 L_{1}=sM^{4} =const
\label{varphi}
\end{equation}
where  $s=\pm 1$ and $M$ is a constant of integration with the
dimension of mass. The appearance of a nonzero integration
constant $sM^{4}$ {\em spontaneously breaks the scale invariance}.

In TMT there is no need\cite{G9}-\cite{G12}  to postulate the
existence of three species for each type of fermions (like three
neutrinos, three charged leptons, etc.). Instead of this we start
from one "primordial" fermion field for each type of leptons and
quarks: one primordial neutral lepton $N$, one primordial charged
lepton $E$, etc.. Splitting of each of them into three generations
occurs as a dynamical effect of TMT in normal particle physics
conditions, that is when fermions are localized (in nuclei, atoms,
etc.) and constitute the regular (visible) matter with energy
density tens orders of magnitude larger then the vacuum energy
density. The crucial role in this effect belongs to the
above-mentioned constraint which dictates the balance (in orders
of magnitude) between the vacuum energy density  and the fermion
energy density. In normal particle physics conditions this balance
may be satisfied if $\zeta$ gets the set of pairs of constant
values $\zeta^{(i)}_{1,2}$ that correspond to two different states
of the each type of primordial fermions ($i=N,E,...$) with {\it
different constant masses}. It turns out that with those constant
values of $\zeta$, the non-canonical fermion contribution to the
energy-momentum tensor disappears and the gravitational equations
of our TMT model are reduced exactly to the Einstein equations in
the corresponding field theory model (i.e. when the scalar field
and massive fermions are sources of gravity). Since the classical
tests of GR deal with matter built of the fermions of the first
generation (with a small touch  of the second generation), one
should identify the states of the primordial fermions obtained as
$\zeta =\zeta_{1,2}^{(i)}$ with the first two generations of the
regular fermions. One can show that the model allows to quantize
the matter fields and provides right flavor properties of the
electroweak interactions, at least for the first two lepton
generations.

It turns out that besides the discussed two solutions for $\zeta$
there is only one more additional possibility to satisfy the
constraint when primordial fermion is in the normal particle
physics conditions and to provide that the non-canonical fermion
contribution to the energy-momentum tensor is much less than the
canonical one. We associate this solution with the third
generation of fermions. (for details see \cite{G9}-\cite{G11}).
The described effect of splitting of the primordial fermions into
three generations in the normal particle physics conditions can be
called "{\em fermion families birth effect}".

Fermion families birth effect (at the normal particle physics
conditions) and reproduction of Einstein equations (as the
fermionic matter source of gravity built of the fermions of the
first two generations) do not exhaust the remarkable features of
the theory. Simultaneously with this the theory automatically
provides an extremely strong suppression of the Yukawa coupling of
the scalar field $\phi$ to the fermions observable in
gravitational experiments. The mechanism by means of which the
model solves the long-range scalar force problem is very unusual:
primordial fermions interact with quintessence-like scalar field
$\phi$, but this interaction practically disappears when
primordial fermions are in the states of the regular fermions
observed in gravitational experiments with visible matter. The
fact that {\em the same condition provides simultaneously both
reproduction of GR and the first two families birth effect} seems
very surprising because we did not make any special assumptions
intended for obtaining these results.

In the fermion vacuum the constraint determines $\zeta$ as the
function of the dilaton $\phi$ (and of the Higgs field if it is
included in the model). If the integration constant is chosen to
be negative ($s=-1$ in Eq.(\ref{varphi})) then the effective
potential of the scalar sector implies a scenario\cite{G7} where
zero vacuum energy is achieved without any fine tuning. This
allows to suggest a new way for {\em resolution of the old
cosmological constant problem}. In models with the Higgs field one
may get such situation multiple times, therefore naturally
obtaining a multiple degenerate vacuum as advocated
in\cite{Nielsen}. If one to choose $s=+1$ then one can  treat the
fermion vacuum  as a model for dark energy in the FRW cosmology of
the late time universe. Assuming that the scalar field
$\phi\rightarrow\infty$ as $t\rightarrow\infty$ we
obtain\cite{G9}-\cite{G12},\cite{G14},\cite{G15} that the
evolution of the late time universe is governed by the sum of the
cosmological constant and the quintessence-like scalar field
$\phi$ with the potential proportional to the integration constant
$M^{4}$ and having the form of a combination of two exponents of
$\phi$. In the more simple model(see \cite{G9},\cite{G12}) where
the potentials for $\phi$ are not included in the original TMT
action at all, the effective $\phi$-potential is generated due to
spontaneous symmetry breaking by Eq.(5) and it has the form of the
exponential potential studied in quintessence
models,\cite{quint1},\cite{quint2}.

Due to the constraint, physics of primordial fermions at energy
densities comparable with the dark (scalar sector) energy density
turns out to be very different from what we know in normal
particle physics.  In this case, the non-canonical contribution
(proportional to $g_{\mu\nu}$) of the primordial fermion into the
energy-momentum tensor can be larger and even much larger than the
canonical one. The theory predicts that in this regime the state
of the primordial fermion is totally different from what we know
in normal particle physics conditions.  For instance, in the FRW
universe, the primordial fermion can participate in the expansion
of the universe by means of changing its own parameters. We call
this effect "Cosmo-Particle Phenomenon"  and refer to such states
as Cosmo-Low Energy Physics (CLEP) states. A possible way to
approach and get up a CLEP state might be spreading of the
non-relativistic neutrino wave packet during its free motion (that
may last a very long time). As the first step in exploration of
Cosmo-Particle Phenomena, we have studied a simplified
cosmological model\cite{G12},\cite{G14},\cite{G15} where the
spatially flat FRW universe  is filled with a homogeneous scalar
field $\phi$ and uniformly distributed {\it non-relativistic
(primordial) neutrinos}. Some of the features of the CLEP state in
this toy model are the following: neutrino mass increases as
$a^{3/2}$ ($a=a(t)$ is the scale factor); its energy density
scales as a sort of dark energy and its equation-of-state
approaches $w=-1$ as $a\rightarrow\infty$; the total energy
density of such universe is less than it would be in the universe
free of fermionic matter at all. The described effect of the
neutrino contribution into the dark energy is much stronger than
the one studied in Ref.\cite{Nelson}.

When including terms quadratic in curvature, these types of models
can be applied not only for the late time universe but also for
the early inflationary epoch. As it has been demonstrated in
Ref.\cite{G13}, a smooth transition between these epochs is
possible in these models.

\section{ Strings, Branes, Horizon and K-K modes.}

With the 2-dimensional version of the measure $\Phi$ we can
construct the world-sheet density
\begin{equation}
\Phi =  \frac{1}{2}\varepsilon^{ab}  \varepsilon_{ij}
\partial_{a} \varphi^{i} \partial_{b} \varphi^{j}.
\label{2-Phi}
\end{equation}
However, a problem appears in the naively generated Polyakov-type
string action $S_{0} = - \frac{1}{2}\int d^{2}\sigma \Phi
\gamma^{ab}
\partial_{a} X^{\mu}\partial_{b} X^{\nu} G_{\mu \nu}$ because the equation
that results from the variation of 
the inverse world-sheet metric 
$\gamma^{ab}$ yields  the unacceptable condition $\Phi \partial_{a}
X^{\mu}\partial_{b} X^{\nu} g_{\mu \nu} = 0$, i.e. vanishing of
the induced metric on the world-sheet. To remedy this situation we
have considered\cite{G16}-\cite{G18} an additional term
$S_{g}=-\int d^{2}\sigma\Phi L$ where  $\sqrt{-\gamma}L$ would be
a total derivative. One can see that without loss of generality,
$L$ may be chosen in the form $\frac
{\varepsilon^{ab}}{\sqrt{-\gamma}} F_{ab}$ where $F_{ab} =
\partial_{a} A_{b} - \partial_{b} A_{a}$, \, $A_{a}(\sigma)$ is an
abelian gauge field on the world sheet of the string.

The action $S_{0}+S_{g}$ is invariant under diffeomorphisms
$\varphi_{i} \rightarrow \varphi_{i}^{'} = \varphi_{i}^{'}
(\varphi_{j})$ in the space of the measure fields (so that $\Phi
\rightarrow \Phi^{'} = J \Phi$) combined with a conformal (Weyl)
transformation of the metric $\gamma_{ab}$: \,  $\gamma_{ab}
\rightarrow \gamma^{'}_{ab} = J \gamma_{ab}$ The combination
$\frac {\varepsilon^{ab}}{\sqrt{-\gamma}} F_{ab} $ is a genuine
scalar. In two dimensions it is proportional to $\sqrt{ F_{ab}
F^{ab}}$.

The equation of motion obtained from the variation of the gauge
field $A_{a}$ is $\varepsilon^{ab}\partial_{a}
(\frac{\Phi}{\sqrt{-\gamma}}) = 0$, which can be integrated to
yield a {\em spontaneously induced string tension}
$T=\frac{\Phi}{\sqrt{-\gamma}}$. The string tension appears here
as an integration constant and does not have to be introduced from
the beginning. The string theory Lagrangian in the modified
measure formalism does not have any fundamental scale associated
with it. The gauge field strength $ F_{ab}$ can be solved from a
fundamental constraint of the theory, which is obtained from the
variation of the action with respect to the measure fields
$\varphi_{j}$ and which requires that $L=M=constant$. Consistency
demands $M=0$ and finally all the equations are the same as those
of standard bosonic string theory.

The described model can be extended\cite{G17},\cite{G18} by
putting point-like charges on the string world-sheet which
interact with the world-sheet gauge field $A_{a}$. Then the
induced tension is not a constant anymore and it suffers
discontinuous jumps at the points where electric charges are
located. The generalization of this model to the non-Abelian gauge
fields is straightforward\cite{G17},\cite{G18} by using $\sqrt{Tr
F_{ab} F^{ab}}$ instead of $\frac
{\varepsilon^{ab}}{\sqrt{-\gamma}} F_{ab} $ (in the non-Abelian
case the latter is not a scalar in the internal space ). In this
case {\em the induced tension is identified as the magnitude of an
effective non-Abelian electric field-strength on the world-sheet
obeying the standard Gauss-low constraint}. As a result, a simple
classical mechanism for confinement via modified-measure "color"
strings has been proposed\cite{G17},\cite{G18} where {\em the
colorlessness of the "hadrons" is an automatic consequence of the
new string dynamics}.

We have studied two types of branes. 
The first one is similar to the Nambu-Goto type of branes well known in
literature, however, with dynamically generated brane tension. The branes
of the second type have totally new features and have no analog in the
literature. In order to construct the bosonic $p$-branes with dynamical
tension, a term of the form $\frac
{\varepsilon^{a_{1}a_{2}...a_{p+1}}}{\sqrt{-\gamma}}
\partial_{[a_{1}}A_{a_{2}...a_{p+1}]}$ has to be added in the Polyakov-type
brane action instead of the $\frac {\varepsilon^{ab}}{\sqrt{-\gamma}} F_{ab}$.
The branes of the second type are constructed\cite{G19},\cite{G20} by adding
term of the form $\sqrt{F_{ab}F^{ab}}$ in order to {\em achieve Weyl-conformal
invariance for any $p$}:
\begin{equation}
S = - \int d^{p+1}\sigma \,\Phi (\varphi)
\Bigl\lbrack \frac{1}{2} \gamma^{ab} \partial_a X^{\mu} \partial_b X^{\nu}
G_{\mu\nu}(X) - \sqrt{F_{ab}(A) F_{cd}(A) \gamma^{ac}\gamma^{bd}}\Bigr\rbrack
\label{WI-brane}
\end{equation}
For any even $p$ the action (\ref{WI-brane}) describes the dynamics of
{\em intrinsically light-like} branes (\textsl{WILL-branes}). 
In particular, for $p=2$ the spherically
symmetric solutions {\em automatically adjust themselves to sit at
the black hole horizon}. This suggests that the second type of
branes can serve as a relevant candidate for realization of the
idea of the black hole membrane paradigm\cite{paradigm} and of the
't Hooft approach\cite{hooft} to description of the degrees of
freedom of the horizon.

In the "Kaluza-Klein" context we have found\cite{G20} solutions
describing \textsl{WILL}-branes wrapped around the internal
(compact) dimensions and moving as a whole with the speed of light
in the non-compact (space-time) dimensions. Although the {\em
WILL}-brane is wrapping the extra dimensions in a topologically
non-trivial way, its modes remain {\em massless} from the
projected $d$-dimensional space-time point of view. This is a
highly non-trivial result since we have here particles (membrane
modes), which acquire in this way non-zero quantum numbers, while
at the same time remaining massless. In contrast, one should
recall that in ordinary Kaluza-Klein theory, non-trivial
dependence on the extra dimensions is possible for point particles
or even standard strings and branes only at a very high energy
cost (either by momentum modes or winding modes), which implies a
very high mass from the projected $D=4$ space-time point of view.

\section{ Braneworld Scenarios.}

A six dimensional braneworld scenario based on a model describing
the interaction of gravity, gauge fields and $3+1$ branes in a
conformally invariant way is described by the action
\begin{equation}
S = \int L\Phi_{(6)} d^{6}x ,\qquad L = -\frac{1}{\kappa_{(6)}}
R^{(6)} + \sqrt{|F_{CD}F^{CD}|}, \label{III-1}
\end{equation}
where $\kappa_{(6)}$ and $R^{(6)}$ are 6-D gravitational constant
and scalar curvature 
and $F_{CD}$ is 6-D gauge field strength. 
The action of this model is defined using a
6-dimensional version $\Phi_{(6)}$ of the measure $\Phi$ (\ref{Phi}).
This allows for the theory to be conformally invariant. In this theory the
branes do not need to be postulated separately. They result here
from delta-function configuration of the gauge fields. As it is
known, $\sqrt{|F_{CD}F^{CD}|}$-gauge theory allows for such type
of extended object solutions \cite{sqrt-1,sqrt-2}. It was shown in
Refs.\cite{sqrt-1,sqrt-2} that in such a model there is no need to fine tune
any bulk cosmological constant or the tension of the two parallel
branes to obtain zero 4-D cosmological constant: the only
solutions are those with zero 4-D cosmological constant. In
contrast, the extra dimensions in these solutions are highly
curved.

\textbf{Acknowledgements.} We gratefully acknowledge support through
the exchange agreement between the Ben-Gurion Univesity of the Negev
and the Bulgarian Academy of Sciences. E.N. and S.P. are partially
supported by Bulgarian NSF grants F-904/99 and F-1412/04.

\end{document}